\documentclass{article}

\usepackage{arxiv}

\usepackage[utf8]{inputenc} % allow utf-8 input
\usepackage[T1]{fontenc}    % use 8-bit T1 fonts
\usepackage{hyperref}       % hyperlinks
\usepackage{url}            % simple URL typesetting
\usepackage{booktabs}       % professional-quality tables
\usepackage{amsfonts}       % blackboard math symbols
\usepackage{nicefrac}       % compact symbols for 1/2, etc.
\usepackage{microtype}      % microtypography
\usepackage{lipsum}
\usepackage{graphicx}
\graphicspath{ {./images/} }
\usepackage{lineno}
\usepackage{subcaption}
\usepackage{amsmath}
\usepackage{xfrac}
\usepackage{setspace}
\usepackage{array}
\usepackage{todonotes}

\DeclareUnicodeCharacter{2212}{-}

\title{\boldmath Lorentz Dissociation of Hydrogen ions in a Cyclotron}

\author{
Hui Wen Koay\\
TRIUMF\\
4004 Wesbrook Mall, \\
Vancouver, B.C. Canada V6T 2A3 \\
 \texttt{hkoay@triumf.ca} 
}

\begin{document}
\onehalfspacing
\maketitle
\begin{abstract}
Stripping extraction of hydrogen molecular ions has gained interest in the cyclotron industry due to its high extraction efficiency. However, the magnetic field could result in undesired Lorentz dissociation of the hydrogen anion/molecular ions during acceleration. Studies of dissociation under electric fields comparable to that of a Lorentz-transformed magnetic field in a typical cyclotron (a few MV/cm) are sparse. Hence, in order to fill in the missing yet crucial information when designing a cyclotron, this work compiles and summarizes the study of Lorentz dissociation of H$^-$, H$_2^+$ and H$_3^+$ for stripping extraction in a cyclotron. 
\end{abstract}

\section{Introduction}
\label{sec:intro}
Stripping extraction takes away one or more electrons from the accelerated particles by passing them through a thin stripper foil located at the desired extraction position, thus also breaking any molecular bonds. As the stripped particles have different charge-to-mass ratio than the accelerated whole particles, the stripped ions are deflected away from the original orbit, possibly leading to nearly 100\% of extraction efficiency \cite{saha2015, mp30}. This feature is appealing especially in the production of a high intensity proton beam. TRIUMF is among the largest cyclotrons that adopted the H$^-$ extraction \cite{yuri2018}, while the DAE$\delta$ALUS Superconducting Ring Cyclotron \cite{daniel2017} and the TR150 cyclotron proposed by Rao et al.\cite{rao2022innovative} are recent projects that propose to extract by stripping of H$_2^+$ and H$_3^+$ ions respectively.

Though stripping ions is promising for clean extraction, there are other issues that are important at high power. One of these, not dealt with here, is the foil lifetime. Another is the so-called Lorentz dissociation of the accelerated ions under the effect of magnetic field. This is a quantum mechanical effect where the bound electron or proton can tunnel out of its potential well because in its own reference frame, the magnetic field is an electric field that tilts the well. Dissociation occurs anywhere along ions' orbits and so they are lost, not extracted and eventually cause activation. Therefore, in order to investigate the impact it has in limiting the beam intensity at different energies, we have studied and compared the Lorentz dissociation of three different hydrogen ions that are good candidates for stripping extraction: H$^-$, H$_2^+$ and H$_3^+$.

\section{Stripping extraction}
\begin{figure}[ht!]
\centering
\begin{subfigure}{0.49\textwidth}
	\centering
	\includegraphics[width=0.95\textwidth]{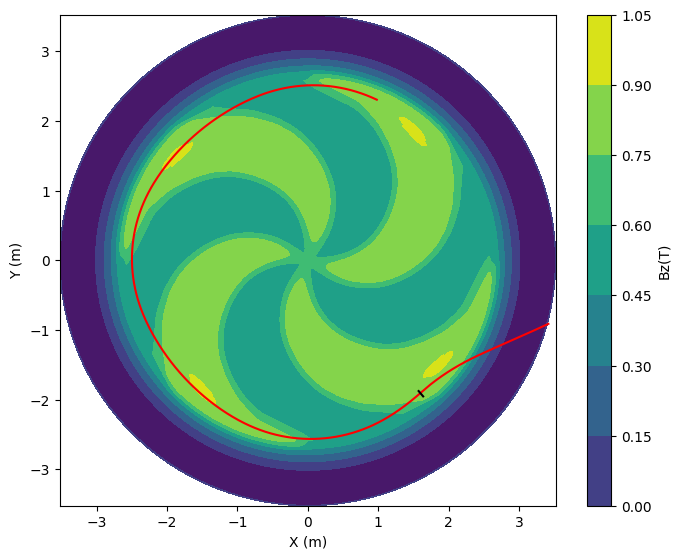}	
	\caption{H$^-$ at $B_0\sim0.7$ T} \label{fig:stripH-}
\end{subfigure}
\begin{subfigure}{0.49\textwidth}
	\centering
	\includegraphics[width=0.95\textwidth]{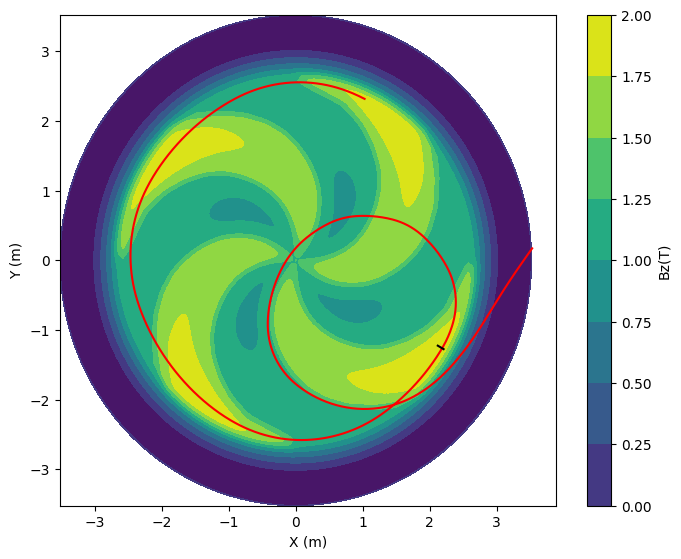}	
	\caption{H$_2^+$ at $B_0\sim1.3$ T} \label{fig:stripH2}
\end{subfigure}

\begin{subfigure}{0.49\textwidth}
	\centering
	\includegraphics[width=0.95\textwidth]{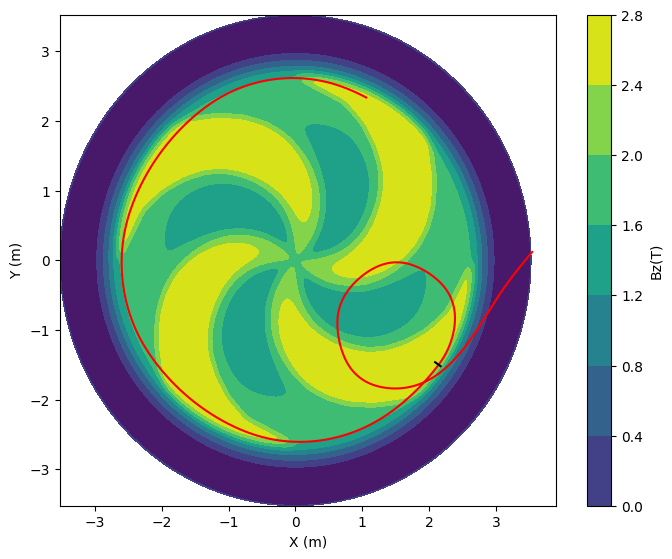}	
	\caption{H$_3^+$ at $B_0\sim1.9$ T} \label{fig:stripH3}
\end{subfigure}
	\caption{Sample trajectories to strip the three different types of H ions (red) in order to obtain 150 MeV/u H$^+$ beams. The stripper foil is indicated by the small black line. Note that $B_0$ is the average central magnetic field at the median plane of the cyclotron.} 
\label{fig:stripH}
\end{figure}

Stripping extraction of the H$^-$ ions is the most common and easiest among the three hydrogen ions. This is because of the exactly opposite charge states of H$^+$, causing them to be naturally deflected out of the cyclotron after being stripped. An example of the trajectory of stripping extraction of H$^-$  is also illustrated in Figure \ref{fig:stripH-}.

On the other hand, unlike the simple opposite trajectory of H$^-$, stripping of H$_n^+$ of the same sign causes the particle to immediately bend inward after stripping. This is due to a smaller mass-to-charge ratio of the stripped particles, resulting in an $n$ times smaller radius. A more complicated beam dynamics is required in order to extract the stripped H$^+$ out of the cyclotron. 
For instance, Figure \ref{fig:stripH2} and \ref{fig:stripH3} show the trajectories of stripping extraction of H$_2^+$ and H$_3^+$ during the last turn. From the figures, the stripped protons have almost half and one-third of the initial radius of H$_2^+$ ions and H$_3^+$ ions respectively. In fact, despite a larger difficulty due to the smallest radius, stripping extraction of H$_3^+$ is better, as the former has a tendency of passing through the central regions, potentially causing undesired beam loss before being extracted from the cyclotron. 

\section{Lorentz dissociation of hydrogen ions}
\subsection{Equivalent electric field}%\todo[color=green!40]{I think this can be shortened considerably, because only magnitude of E' matters. Also, I think it's wrong since $E_{\parallel}'=E_{\parallel}$. See \href{https://en.wikipedia.org/wiki/Classical_electromagnetism_and_special_relativity\#The_E_and_B_fields}{here}. Further, I removed the section on $\gamma$ depending only on energy per proton, and just said it in words.}
In the particle's rest frame, the electric field components parallel to and perpendicular to the motion are:
%From Lorentz transformation, the equivalent electric field in the rest frame of a charged particle moving with velocity $\textbf{v}$ is given by
\begin{equation}
%E_{\parallel}'=\frac{( \textbf{E}+\textbf{v}\times \textbf{B})_{\parallel}}{\sqrt{1-\beta^2}} 
\mathcal{E}_{\parallel}=\textbf{E}_{\parallel},\ \mbox{ and }\ \mathcal{E}_{\bot}=\frac{( \textbf{E}+\textbf{v}\times \textbf{B})_{\bot}}{\sqrt{1-\beta^2}} \label{eq:lor1}
\end{equation}
where \textbf{E} and \textbf{B} are the external electric and magnetic field respectively; $\beta$ is the ratio of the particle's speed, $v$, to the speed of light, $c$. Whether H$^-$, H$_2^+$ or H$_3^+$, for a given energy per proton, all have the same $\beta$ and $\gamma=\frac{1}{\sqrt{1-\beta^2}}$.

As \textbf{v} and \textbf{B} are at right angles, and $vB_z\gg \mathcal{E}$, we have 
%Considering only the effect in radial direction, equation \ref{eq:lor1} can be simplified as:
%\begin{equation}
%E_{r}'=\frac{E_r+vB_z}{\sqrt{1-\beta^2}}  \nonumber \\
%=\gamma( E_r+vB_z)
%\label{eq:lor2}
%\end{equation}
%As contribution of $B_z>>E_r$ in a cyclotron, we can treat equation \ref{eq:lor2} as follows:
\begin{equation}
\mathcal{E}=\gamma \beta c B_z  \cong (3\text{ MV/cm}) \gamma \beta (B_z/1\,\rm T) \label{eq:lor3}
\end{equation}
%For convenience, $E_r'$ will be expressed simply as $\mathcal{E}$ in the following discussion of this work. 

% \begin{align*}
% \text{As }\gamma=\frac{E_0+T}{E_0}\\
% =1+\frac{T}{E_0} \\
% \gamma  & \approx 1 + \frac{A (T/n)}{A ( E_{H+})} \\
%   & \approx 1 + \frac{T/n}{E_{H+}} 
% \end{align*}
% given $T$ as the total kinetic energy of ions; $A$ as the atomic mass number; $E_0$ as the rest mass energy of the ion; $E_{H+}$ as the proton mass energy. If the final kinetic energy per nucleon, $T/n$, remains unchanged, $\gamma \beta$ and the equivalent electric field,  $\mathcal{E}$, are similar for H$^-$, H$_2^+$ and H$_3^+$ at the same $B_z$.

\begin{figure}[ht!]
	\centering
		\includegraphics[width=0.5\textwidth]{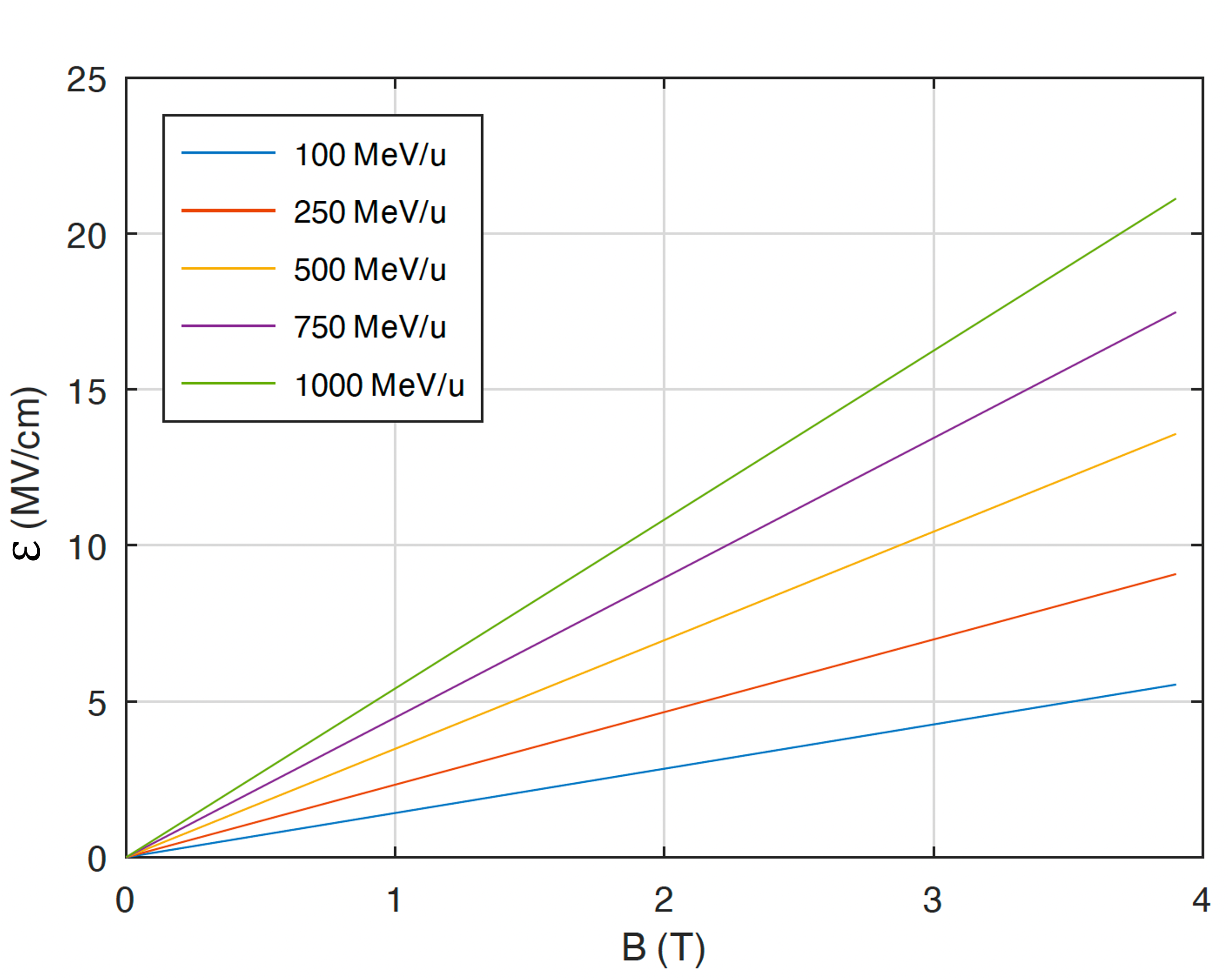}	
		\caption{The equivalent electric field  $\mathcal{E}$ of hydrogen ions} \label{fig:Equiv}
\end{figure} 

\subsection{\texorpdfstring{H$^-$}{H-}}
H$^{-}$ is extracted by H$^{-}\rightarrow$ H$^{+}$ + 2e$^{-}$. Despite the simplicity and efficiency of this method, the small binding energy (about 0.7 eV) of the second electron in the H$^-$ ions is a potential source of beam loss during acceleration. The probability for this occurrence is characterized by a time constant $\tau$: the probability of surviving a time $t$ is $e^{-t/\tau}$. Using a WKB method, one can show that $\tau$ has the following form \cite{darewych1963mean}:
\begin{equation}
    \tau= \frac{A_1}{\mathcal{E}} \exp{\frac{A_2}{\mathcal{E}}} 
\end{equation}
The constants $A_1$ and $A_2$ could in principle by calculated from a full quantum-mechanical treatment, but have been fitted from experimental results \cite{keating1995electric}. They are: $A_1 = 3.07 \times 10^{-6}$ V-s/m and $A_2 = 4.414 \times 10^9$ V/m respectively. Due to the popularity of stripping extraction of H$^-$, this effect is well studied and documented by many past researchers \cite{stinson1969,keating1995electric}.

\begin{figure}[htb!]
\centering
	\includegraphics[width=0.8\textwidth]{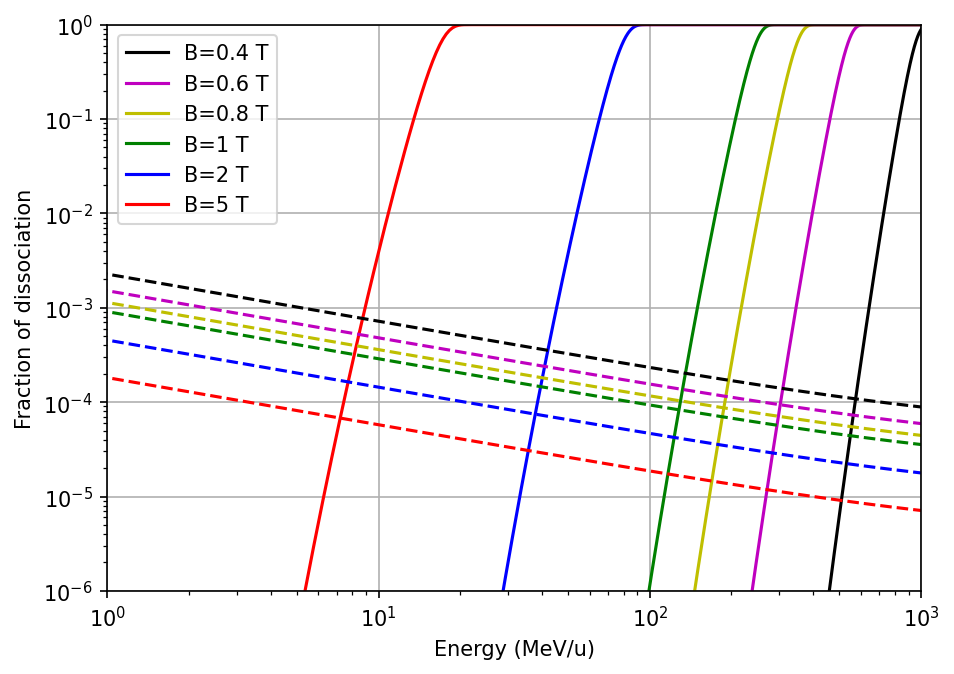}	
	\caption{Integrated fractional dissociation of H$^-$ as a function of energy for various $B$. Note that this calculation assumed a constant $B$ and the energy gain per turn is 0.48 MeV. However since the loss as such a steep function of $B$, this can be taken as the peak $B$ for cyclotrons with flutter. The dashed lines show the maximum allowed fraction for 1\,W/m lost on the vacuum chamber wall assuming 1 mA beam. The crossing points between the dashed and solid lines mark the maximum operating energy at respective $B$.}
	\label{fig:lorentzH-}
\end{figure}

To convert survival fraction $f$ per time to loss per distance ($s$) travelled requires an extra factor of $\gamma$ for time dilation:
\begin{equation}
%F&=1-f \\
\frac{df}{ds}=-\frac{f}{\gamma \beta c \tau} \label{eq:lorentz1} 
%\ln \left( \frac{f_2}{f_1} \right) &= - \frac{s_2-s_1}{\gamma \beta c \tau}  \nonumber \\
%\therefore  f_2&= f_1 \exp(-\frac{\Delta s}{\gamma \beta c \tau}) \\
%\tau&= \frac{A_1}{\mathcal{E}} \exp{\frac{A_2}{\mathcal{E}}}  \nonumber
\end{equation}
%\noindent where $\tau$ is the rest-frame life time of the H$^-$ ion; $f$ is the H$^-$ fraction surviving in the beam and $s$ is the length along the beam path. 
and as $\tau$ is a function of both $\beta\gamma$ and $B$, this is to be integrated along a particle's path. In the following estimation, we assume $B$ is a constant, i.e.\ omitting any flutter and the energy gain per turn is 0.48 MeV. The integrated fractional loss $F=1-f$ is shown in Figure \ref{fig:lorentzH-}.

The loss limit of a proton beam is usually taken to be $\sim1$\,W/m for hands-on maintenance \cite{triumfFYP,EuCARD-2}. Losses due to Lorentz stripping will be spread along the inside surface of the outer wall of the vacuum chamber. Taking the circumference to be $2\pi\hat{R}$ and $\hat{R}=\frac{\hat{P}}{eB}$, with the top momentum $\hat{P}$, the power loss limit $\mathcal{P}_m$ can be estimated. The ratio of $\frac{\mathcal{P}_m}{\text{Beam power}}$ of a 1 mA beam at respective $B$ are also shown by dashed lines in Fig.\,\ref{fig:lorentzH-}. The crossovers of dashed and solid curves, which indicate largest energy for given field, trend upward, reflecting the fact that higher energy requires larger machines, spreading loss over a larger area.

From Fig.\,\ref{fig:lorentzH-}, $F$ of H$^-$ particles accelerated up to 1\,GeV is 100\% for any $B>0.4$\,T. Taking a beam current of 1 mA, the maximum permissible beam energy at a low $B$ of 0.4\,T is 570 MeV; at a very large radius of $r=10$\,m. Hence, any acceleration at high energy is uneconomic, as the average magnetic field will be too low to achieve the desired beam loss, and the machine has to be extremely large to accommodate such a low magnetic field.

To date, only the TRIUMF cyclotron accelerates H$^-$ to the high energy region ($> 400$ MeV); the higher the energy, the lower the magnetic field must be to keep Lorentz stripping low. The TRIUMF cyclotron has a low average $B<0.46$\,T and a large extraction radius of $\sim9$\,m. In order to achieve a competitive level of meson production, the H$^-$ beam is accelerated up to 500\,MeV. To limit the Lorentz loss to a maximum beam loss of 5.2\% \cite{yuri2014} required the peak B field to be no more than 0.56 T and this reduced the flutter at the highest energy to $\sim 0.07$. To maintain vertical stability, the reduced flutter had to be compensated with increased spiral angle of the sectors. This resulted in a rather extreme $72^\circ$ spiral angle in the higher energy region, and made it impossible to place resonators between sectors. This in turn resulted in quarter-wave resonators mounted horizontally inside the magnet gap, and an enlarged the magnet gap of 0.5\,m.

\subsection{\texorpdfstring{H$_2^+$}{H2+}}
H$_2^+$ is a diatomic ion with an equilibrium bond distance of about 1.06\,\AA. The binding energy of H$_2^+$ is about 2.7 eV, which is 3.6 times larger than the binding energy of H$^-$. Unlike H$^-$ ions, a complete break up of H$_2^+$ into its constituent particles due to the electron tunneling through the deformed electronic potential well is not the main mode of dissociation. Instead, the presence of $\mathcal{E}$ lowers the asymptotic nuclear potential of the lower electronic state  \cite[Fig.3]{hiskes1961}. This increases the chance of H$_2^+$ at high-lying vibrational states to become unstable and a proton to ``leak" from the potential well. The molecular ion then disintegrates into a proton and a hydrogen atom and this is called the pre-dissociation; the main dissociation mode of H$_2^+$. 

Past experiments had shown evidence of Lorentz dissociation of  H$_2^+$ at a much lower equivalent electric field ($\mathcal{E}<1$ MV/cm)\cite{herman1963,riviere1960}. Due to the lack of experimental data at a higher field (typical range used in a cyclotron: $\mathcal{E}>1$ MV/cm), some theoretical models have to be used as the preliminary tools to estimate the Lorentz dissociation of H$_2^+$ in a cyclotron.

\begin{figure}[ht!]
\centering
	\includegraphics[width=0.7\textwidth]{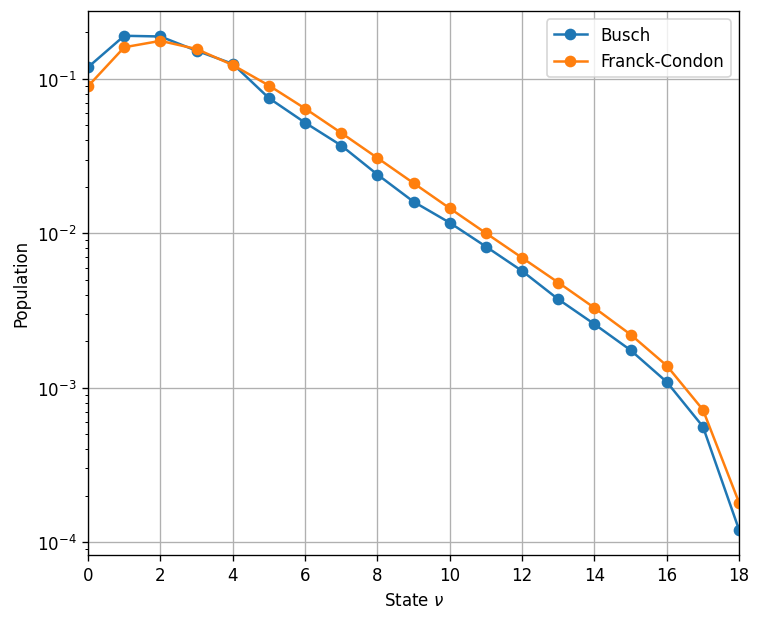}
	\caption{State populations of H$_2^+$. Blue line shows results from Busch \cite{busch1972} while orange line shows the one determine using the Franck-Condon model\cite{franck1966}.}
	\label{fig:busch}
\end{figure}

The ionic lifetime of each vibrational $\nu$ state ($J=0$) at different electric fields can be obtained from Hiskes' calculation \cite{hiskes1961}. The lifetimes of high $\nu$ states ($\sim 10^{-8}$\,s) are generally comparable to the revolution period of H$_2^+$ in a cyclotron, i.e.\ the ions at high $\nu$ states will dissociate completely within a turn of revolution. Therefore, instead of the lifetime, it is the state population that limits the fraction of dissociation. Unlike H$^-$ that has only one bound state, H$_2^+$ has many populated bound states that highly depend on the initial conditions of the source. Many studies had shown consistently that more than 90\% of the population lie at $\nu<12$ \cite{brenton1984,busch1972,ozenne1976,franck1966}. Among them, the most important work was done by Busch, as a comparison between calculation and experiment was made \cite{busch1972}. The distribution of the state population by Busch is also given in Fig.\,\ref{fig:busch} .

\begin{figure}[ht!]
\centering
\begin{subfigure}{0.8\textwidth}
	\centering
	\includegraphics[width=0.95\textwidth]{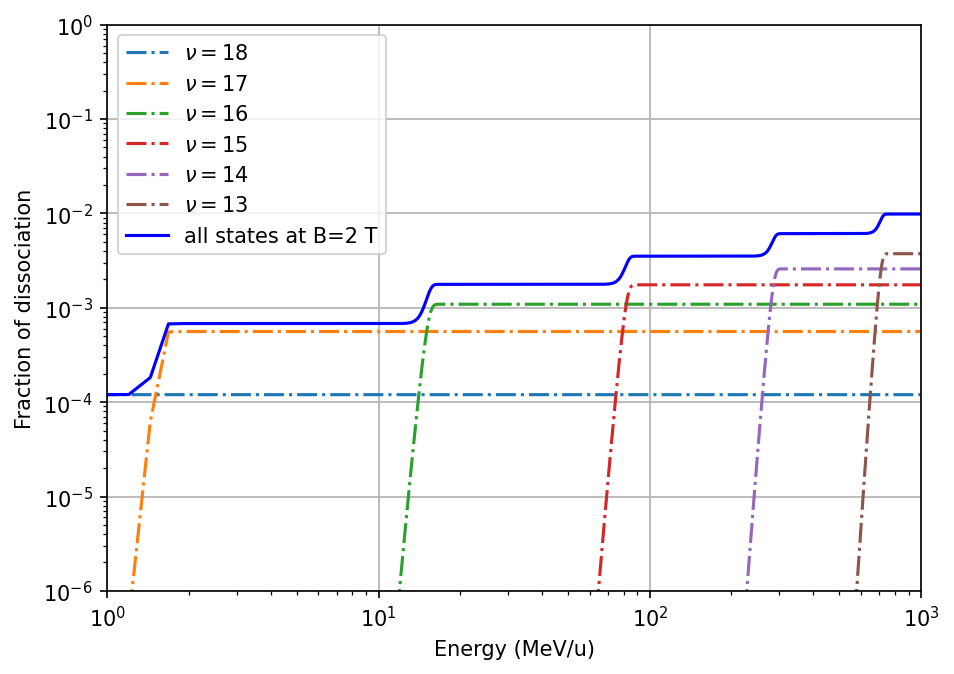}	
	\caption{Dissociation of unstable $\nu$ states at B=2 T} \label{fig:h2_dissociate_a}
\end{subfigure}

\begin{subfigure}{0.8\textwidth}
	\centering
	\includegraphics[width=0.95\textwidth]{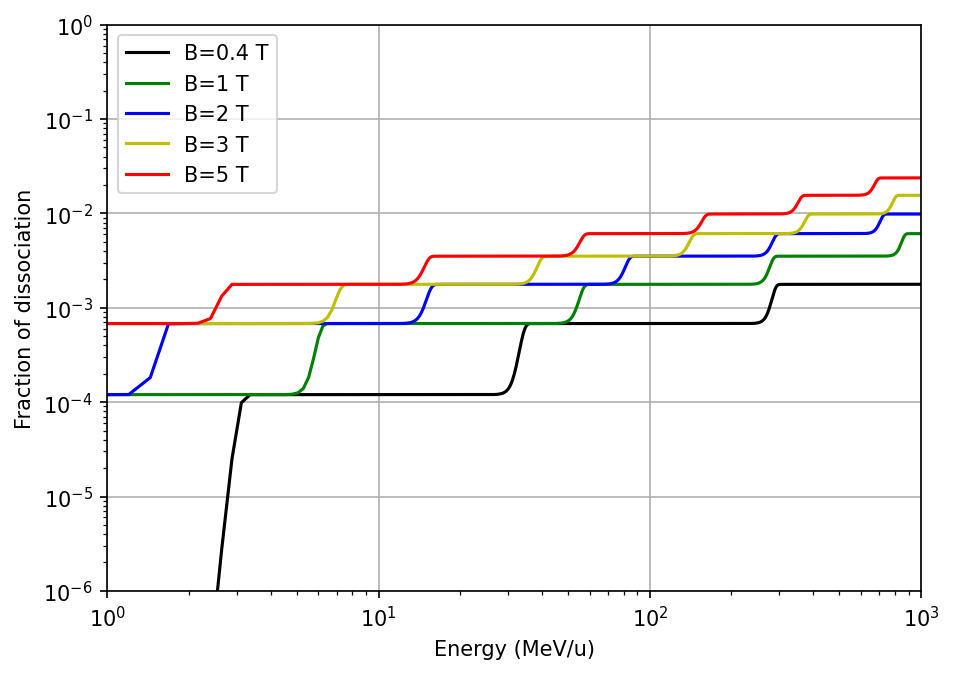}	
	\caption{Total dissociation of all state at B=1, 2, 3 and 5 T} \label{fig:h2_dissociate_b}
\end{subfigure}
	\caption{A summary plot of the dissociated fraction of H$_2^+$. (a) shows the contribution from individual unstable state when $B=2$\,T, while (b) shows the total dissociation from all $\nu$ states when $B=0.4, 1, 2, 3$ and 5\,T respectively.  Each plateau (dashed-dot line) in (a) indicates the maximum dissociated fraction at each $\nu$ state. This corresponds to the population of states in \cite{busch1972}. The blue line is the summation of all vibrational states at $B=2$\,T. As $B$ increases from 1 to 5 T, the total dissociated fraction also increases up to $\sim 2\%$.}
	\label{fig:h2_dissociate}
\end{figure}

If we assume the state populations $p_{\nu}$ are distributed as Busch and the lifetime prediction is as given by Hiskes, the integrated fractional loss $F=\sum_{\nu}p_{\nu}(1-f_{\nu})$ of H$_2^+$ at different $B$ fields can be estimated. Similar to the calculation done for H$^-$, flutter is omitted and the energy gain per turn is taken as 0.48 MeV. This result is summarized in Fig.\,\ref{fig:h2_dissociate}. 
%The 1\,W/m power loss limit per mA for hands-on-maintenance at $B=2$\,T is also shown by a blue dashed line in Fig.\,\ref{fig:h2_dissociate}. 

At $B=2$\,T, the equivalent electric field for 100\,MeV/u of H$_2^+$ is 2.8\,MV/cm. At this $\mathcal{E}$, 0.35\% of ions lying at $\nu \geq 15$ will dissociate according to Fig.\ \ref{fig:h2_dissociate_a}. As the energy or magnetic field increases, $\mathcal{E}$ increases, and so does the number of populated $\nu$ states that are prone to dissociation. For example, at energy of 1 GeV/u, ions lying at $\nu \geq 13$ and at $\nu \geq 11$ will dissociate when $B=2$\,T and 5\,T respectively. This amounts to a total dissociated fraction of about 1\% and 2\% respectively. If we take the maximum permissible power loss of 1\,W/m for hands-on-maintenance, the maximum current allowed at $B=2$\,T and 5\,T for acceleration up to 1\,GeV/u are merely 1.8 and 0.7\,$\mu$A. If acceleration at a higher current (say 1\,mA) is desired, the H$_2^+$ beam has to be cooled or state-selected so that more than 99.9\% lie at $\nu \leq 11$ \cite{sen1987,schmidt2020}.

%A dissociation of about 0.35\% of a 100 MeV/u H$_2^+$ corresponds to 200 W for a 0.28 mA beam. This is close to the practical limit from many experiences of other laboratories to allow a routine hands-on maintenance \cite{yang2013}.

\subsection{\texorpdfstring{H$_3^+$}{H3+}}
Ever since the first discovery of H$_3^+$  by J.J.\,Thomson in 1911, this poly-atomic molecule has gained much research interest due to its high abundance in both laboratory-scale hydrogen discharges and interstellar space \cite{thomson1911,oka2006}. Its equilateral triangle structure with an equilibrium distance of $\sim0.9$ \AA\ provides a good symmetry and stability to the molecular ions \cite{christ1964,gaillard1978}. The binding energy (dissociation energy) of H$_3^+$ is about 4.5 eV\cite{carrington1984}, which is about 2 times larger than H$_2^+$, and is thus the most stable among the hydrogen ions discussed in this work. 
%Despite its stability, previous works had also recorded the Lorentz dissociation of H$_3^+$ \cite{haste1965}.

\begin{figure}[ht!]
\centering
	\includegraphics[angle=0,width=0.8\textwidth]{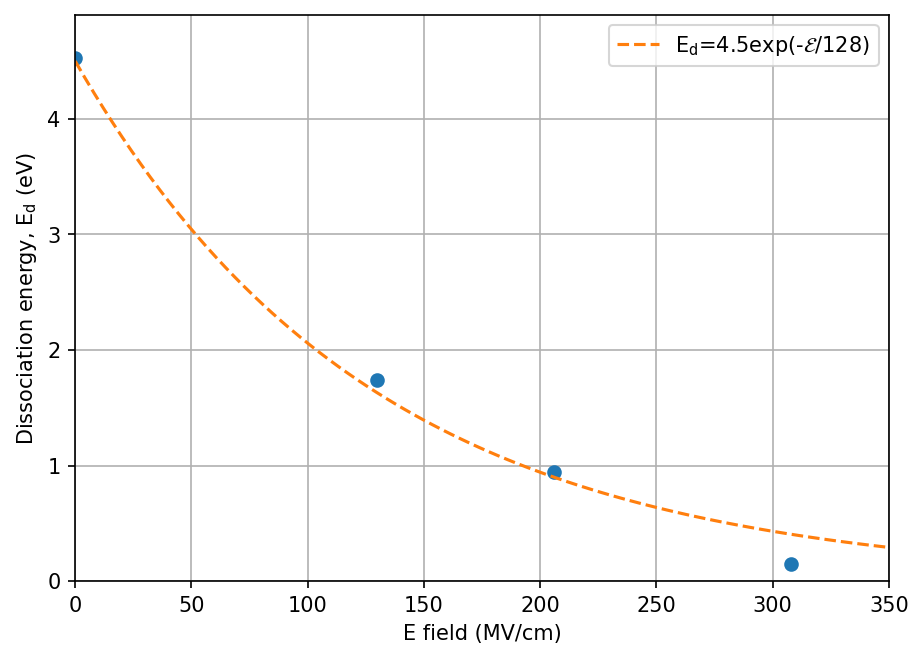}	
	\caption{Fitted function of electric field and dissociation energy for triangular-shaped H$_3^+$ using equation \ref{eq:Ed_h3}.} \label{fig:fitted_h3}
\end{figure}

So far, only very few works studied directly the effect of external field on the dissociation of H$_3^+$. This is due to the complex dynamical structure of the non-linear tri-atomic molecule \cite{haste1965,potentialH3+,medel2013}. Reckz\"ugel et al.\ are among the few who had looked into this for the case of a linear and triangular H$_3^+$ \cite{potentialH3+}. Fig.\,5 in \cite{potentialH3+} shows the change of the potential energy surface of a triangular H$_3^+$ as the external electric field increases. A higher electric field lowers the dissociation energy barrier, causing the ions to disintegrate more easily into a proton and a hydrogen molecule. The relationship between the dissociation energy ($E_{d}$) and $\mathcal{E}$ field (in MV/cm) is extracted from  \cite{potentialH3+} and plotted in Fig.\,\ref{fig:fitted_h3} with a fitted function given as follows:
\begin{equation}
E_{d}=(4.5\,{\rm eV}) \exp\left[-\dfrac{\mathcal{E}}{128\,{\rm MV/cm}}\right] \label{eq:Ed_h3}
\end{equation}
An implication of eqn. \ref{eq:Ed_h3} is that it requires a field of at least $\sim220$\,MV/cm (or $\beta\gamma B\sim70$\,Tesla in the lab frame) to bring the molecule down to the stability level of the H$^-$. But as with the H$_2^+$ case, it is excited states and their population that set the limit. 

H$_3^+$ has two main vibrational modes: the symmetric ``breathing" mode and the asymmetric ``bending" mode. Only the symmetric breathing mode exists when $J=0$; whereas both modes exist when $J>0$. In fact, there are over hundreds of bound rotation-vibrational (ro-vibrational) states with non-zero quantum numbers in these two modes \cite{munro2005}. The full population of all these states with transition time is not easy to determine. V.G.\,Anichich had computed a simpler estimation of the state populations of only the symmetric mode with a quantum number $\nu$ forming from \cite{anicich1984}
\begin{equation}
\mathrm{H}_2^+(\nu_i)+\mathrm{H}_2(\nu_0)\rightarrow \mathrm{H}_3^+(\nu)+\mathrm{H(1s) } \label{reac:h3+}
\end{equation}
by using a cold H$_2^+$ beam (initial $\nu_0=\nu_i=0$). %A more detailed results of the population of excited states using different one- and two-anharmonic models can be found in \cite{anicich1984}. 
As there is no work done so far to determine the lifetime of $\nu$ states at various $\mathcal{E}$, here we assume that it is short as compared to the acceleration period (as in the case of H$_2^+$). The maximum dissociation at a particular $\mathcal{E}$ can then be estimated by utilizing the state populations and eqn. \ref{eq:Ed_h3}. Fig.\,\ref{fig:Ed_pop} shows the state population as distributed in the one-harmonic model from \cite{anicich1984}.

%Similar to the case of H$_2^+$, assuming the state population is as distributed in the one-harmonic model from \cite{anicich1984}, the integrated dissociated fraction $F$ of H$_3^+$ at different magnetic fields can be estimated. Note that this calculation utilizes the lifetime formulation of a heteronuclear-molecule (H$_2+$H$^+$) in \cite{hiskes1961}. The results are shown in Fig.\,\ref{fig:Ed_h3}.

\iffalse
\begin{figure}[ht!]
\centering
	\includegraphics[width=0.8\textwidth]{graphics/h3_B_loglog.png}	
	\caption{The maximum dissociated fraction from the population of unstable states of H$_3^+$ at B=2, 3 and 5 T respectively. Note that this plot utilizes the formulation for a heteronuclear-molecule (H$_2+$H$^+$) in \cite{hiskes1961} and the population of state from the one-harmonic model in \cite{anicich1984}. Each plateau corresponds to a complete dissociation of all the particles at each $\nu$ state. } \label{fig:Ed_h3}
\end{figure} 
\fi

\begin{figure}[ht!]
\centering
	\includegraphics[width=0.8\textwidth]{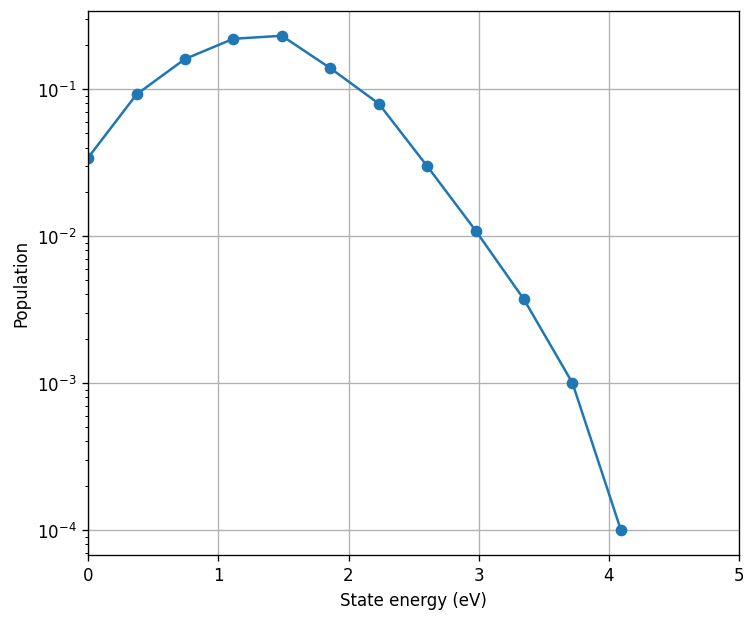}	
	\caption{The population of state from the one-harmonic model in \cite{anicich1984}. Each point corresponds to a bound $\nu$ state from $\nu=0$ at the left to $\nu=11$ at the right.} \label{fig:Ed_pop}
\end{figure} 

Taking H$_3^+$ of energy 1 GeV/u under a constant B field of 3 T, the equivalent $\mathcal{E}$ is about 16.3 MV/cm. This corresponds to a dissociation energy of about 4.0 eV and a dissociation of only the highest state with a population of about 0.01\% from Fig.\,\ref{fig:Ed_pop}. The result is similar even if the population of the more detailed two-anharmonic model from \cite{anicich1984} were used. If we assume a maximum beam loss of 1\,W/m, acceleration of 1\,mA of H$_3^+$ at a high energy of 1 GeV/u is possible at low $B=1.7$\,T. Note that this is about two times the maximum energy per nucleon attainable by H$^-$ at the same radius of 10 m and the same beam current of 1\,mA. Therefore, if the state population of an actual H$_3^+$ beam can be controlled  or cooled \cite{sen1986,urbain2019} so that it is similar to the one adopted here ($>99.9\%$ lying at states with dissociation energy $>4.0$\, eV), we can infer that the effect of external Lorentz field on the beam loss of H$_3^+$ is very minimal at high-energy ($>500$ MeV) extraction.

%\newpage
\section{Comparisons of H ions}
The particle's mass $m_0$, radius $\rho$ and magnetic field $B$ are related to each other by
\begin{align*}
B\rho = m_0 c \gamma \beta   
=  A \gamma \beta (3.1 \text{Tm})
\end{align*}
Therefore, under the same energy per nucleon (same $\gamma \beta$), in order to keep the radius constant, a larger magnetic field is necessary to accelerate hydrogen ions with a higher molecular state. It is important to take into account of these physical factors in addition to the Lorentz dissociation when choosing the most suitable type of ion to achieve the desired beam power. Table \ref{tab:H3_compare} summarizes important parameters to be considered when designing a cyclotron at three different energy regions for the three hydrogen ions discussed.

\begin{table} [ht!]
\centering
\begin{tabular}{p{0.4\textwidth}>{\centering}p{0.1\textwidth}>{\centering}p{0.1\textwidth}>{\centering\arraybackslash}p{0.1\textwidth}}
\hline  \hline  
& H$^-$ & H$_2^+$ &H$_3^+$\\
\hline
Energy (MeV/u) & \multicolumn{3}{c}{100}\\
Momentum, $\beta \gamma$ & \multicolumn{3}{c}{0.47}\\
Extraction radius, $\rho$ (m) & \multicolumn{3}{c}{1.5}\\
Ave. B field, $B_0$ (T) &1 & 2 & 3\\
Peak B field (T) & 1.5 & 3 & 4.5 \\
Max. Lorentz dissociation (\%) & 0.4 & 0.4 & $<0.01$\\
\hline
Energy (MeV/u) & \multicolumn{3}{c}{500}\\
Momentum, $\beta \gamma$ & \multicolumn{3}{c}{1.2}\\
Extraction radius, $\rho$ (m) & \multicolumn{3}{c}{3.64}\\
Ave. B field, $B_0$ (T) &1 & 2 & 3\\
Peak B field (T) & 1.5 & 3 & 4.5 \\
Max. Lorentz dissociation (\%) & 100 & 1 & 0.01\\
\hline
Energy (MeV/u) & \multicolumn{3}{c}{1000}\\
Momentum, $\beta \gamma$ & \multicolumn{3}{c}{1.8}\\
Extraction radius, $\rho$ (m) & \multicolumn{3}{c}{5.65}\\
Ave. B field, $B_0$ (T) &1 & 2 & 3\\
Peak B field (T) & 1.5 & 3 & 4.5 \\
Max. Lorentz dissociation (\%) & 100 & 1.5 & 0.1
\\ \hline \hline  
\end{tabular}
\caption{Comparison of Lorentz dissociation of the three hydrogen ions discussed in this work. Note that the peak B field is obtained by assuming a flutter component of $f_N=0.5$ for $B=B_0(1+f_N\cos{N \theta})$.} \label{tab:H3_compare}
\end{table}

As for the calculation of the maximum Lorentz dissociation of H$^-$ and H$_2^+$ in Table \ref{tab:H3_compare}, a flutter component of $f_N=0.5$ for $B=B_0(1+f_N\cos{N\theta})$ with $N=4$ were included and the Lorentz loss was integrated over all turns up to the final energy using equation \ref{eq:lorentz1}. On the other hand, the dissociation of H$_3^+$ is merely the dissociation of states at the peak field, assuming the population of states remain unchanged throughout the acceleration.

The comparison shows that the best candidate for acceleration up to 100 MeV/u is H$^-$. This is due to its small effect of Lorentz dissociation and the highest cost efficiency with the least B field at the same extraction radius. However, as the beam energy increases, the significant Lorentz dissociation of H$^-$ outweighs this advantage. Thus, when the particle energy is more than 100 MeV/u, acceleration of H$_2^+$ or H$_3^+$ are the better options. The overall stability of H$_3^+$ is, however, slightly better than H$_2^+$, despite the requirement of magnetic field is about 1.5 times higher. This is consistent with the study in \cite{reckzugel1994}. At a very high energy up to 1 GeV, H$_3^+$ shall be the best option due to its lowest Lorentz dissociation.

\section{Conclusions and Prospects}
This work compiled the estimation of Lorentz dissociation for three different types of hydrogen ions at different energy ranges. Overall, the order of stability goes from H$_3^+>$H$_2^+>$H$^-$ at energy greater than 100 MeV\u. If the higher magnetic field is not an issue, H$_3^+$ is the best candidate for acceleration at energy greater than 100 MeV/u while H$^-$ is the most cost effective for low-energy acceleration up to 100 MeV/u. In general, a larger machine is more conducive for stripping extraction at a high power for two reasons: (1) less Lorentz dissociation as $B$ is reduced, (2) a larger power loss permissible for hands-on maintenance.

Nevertheless, the estimations given in this work, especially for H$_2^+$ and H$_3^+$, are very rough and many factors have been omitted. In fact, higher excited states with $J\neq0$ exist and the total number of populated excited states may exceed the reported values from the literature used in this work. As for H$_3^+$, detailed study of the dependence of ionic lifetime of each state on the external field is lacking. Besides, the effect of magnetic field and electric field on the accelerated ions might not be entirely similar. In real practice, the Lorentz dissociation of H$_2^+$ and H$_3^+$ are more complex and it could vary by more than a factor 10, as it highly depends on the initial beam condition \cite{riviere1960,herman1963}. Some facilities have shown the feasibility of generating a high-intensity beam from an ion source for the production of hydrogen ions at different molecular states \cite{moak1959,wu2019}. Therefore, verification of Lorentz dissociation using real hydrogen ions from an ion source should be realistically feasible in the coming future. This shall be the most important work to be done before H$_2^+$ and H$_3^+$ ions can be fully implemented at a higher beam power.

\bibliographystyle{unsrt}
\bibliography{Database}
\end{document}